\documentclass[manuscript,screen,nonacm]{acmart}

\usepackage{multirow}

\AtBeginDocument{%
  \providecommand\BibTeX{{%
    \normalfont B\kern-0.5em{\scshape i\kern-0.25em b}\kern-0.8em\TeX}}}

\setcopyright{acmcopyright}
\acmYear{2024}
\setcopyright{acmlicensed}\acmConference[CACML 2024]{2024 3rd Asia Conference on Algorithms, Computing and Machine Learning}{March 22--24, 2024}{Shanghai, China}
\acmBooktitle{2024 3rd Asia Conference on Algorithms, Computing and Machine Learning (CACML 2024), March 22--24, 2024, Shanghai, China}
\acmDOI{10.1145/3654823.3654895}
\acmISBN{979-8-4007-1641-6/24/03}

\begin{document}

\title[Parameterizations for Gradient-based MCMC on the Stiefel Manifold]{Parameterizations for Gradient-based Markov Chain Monte Carlo on the
Stiefel Manifold: A Comparative Study}

\author{Masahiro Tanaka}
\email{gspddlnit45@toki.waseda.jp}
\orcid{0000-0002-2269-8468}
\affiliation{%
  \institution{Faculty of Economics, Fukuoka University}
  \streetaddress{8-19-1 Nanakuma, Jonan}
  \city{Fukuoka}
  \state{Fukuoka}
  \country{Japan}
  \postcode{814-0180}
}


\begin{abstract}
Orthogonal matrices play an important role in probability and statistics, particularly in high-dimensional statistical models. Parameterizing these models using orthogonal matrices facilitates dimension reduction and parameter identification. However, establishing the theoretical validity of statistical inference in these models from a frequentist perspective is challenging, leading to a preference for Bayesian approaches because of their ability to offer consistent uncertainty quantification. Markov chain Monte Carlo methods are commonly used for numerical approximation of posterior distributions, and sampling on the Stiefel manifold, which comprises orthogonal matrices, poses significant difficulties. While various strategies have been proposed for this purpose, gradient-based Markov chain Monte Carlo with parameterizations is the most efficient. However, a comprehensive comparison of these parameterizations is lacking in the existing literature. This study aims to address this gap by evaluating numerical efficiency of the four alternative parameterizations of orthogonal matrices under equivalent conditions. The evaluation was conducted for four problems. The results suggest that polar expansion parameterization is the most efficient, particularly for the high-dimensional and complex problems. However, all parameterizations exhibit limitations in significantly high-dimensional or difficult tasks, emphasizing the need for further advancements in sampling methods for orthogonal matrices.
\end{abstract}

\ccsdesc[500]{Mathematics of computing~Probabilistic algorithms}
\ccsdesc[500]{Mathematics of computing~Probabilistic inference problems}
\ccsdesc[500]{Computing methodologies~Uncertainty quantification}
\ccsdesc[500]{Mathematics of computing~Bayesian computation}

\keywords{Markov chain Monte Carlo, orthogonal matrix, Stiefel manifold, parametrization}

\maketitle

\section{Introduction}

Orthogonal matrices play essential roles in probability and statistics.
Against the backdrop of the prosperity of big data, high-dimensional
statistical models have gain increasing attention, many of which
are parameterized in terms of orthogonal matrices for 
dimensionality reduction and parameter identification, for example, principal
component analysis, factor analysis, matrix completion, reduced rank
regression, and models for network data \citep{Shi2017,Du2023}. It
is difficult to establish the theoretical validity of statistical
inference in this class of models from a frequentist perspective. Therefore,
a Bayesian approach to inference in models parameterized with
orthogonal matrices is a natural choice because it can provide 
consistent uncertainty quantification, that is, interval estimates, and
hypothesis testing.

Because a posterior distribution is generally intractable, it is necessary
to obtain its numerical approximation. The standard approach is Markov
chain Monte Carlo (MCMC) \citep{Liu2004}, which can provide asymptotically
exact approximation of the target distribution, unlike other methods
such as the Laplace approximation and variational Bayes \citep{Bishop2006}.
However, sampling on the Stiefel manifold, the set of orthogonal
matrices, is notoriously difficult.

Several strategies have been proposed for sampling on the Stiefel manifold.
Hoff \citep{Hoff2009} proposed a rejection-sampling algorithm which
is employed in \citep{Lin2017,Pal2020}. Byrne and Girolami \citep{Byrne2013}
introduced the geodesic Monte Carlo (GMC). Currently, the most efficient approach
is to parameterize orthogonal matrices with unconstrained parameters and simulate from the unconstrained
space using a no-U-turn sampler (NUTS) \citep{Hoffman2014},
an adaptive version of the Hamiltonian Monte Carlo method \citep{Duane1987,Neal2011}.
For this purpose, four alternative parameterizations have been considered:
Householder parameterization \citep{Nirwan2019}, Cayley transform
\citep{Jauch2020}, Givens representation \citep{Pourzanjani2021},
and polar expansion \citep{Jauch2021}. 

There has been no thorough comparison between the parameterizations and
the existing literature only partially provides information about
their relative advantages. In \citep{Jauch2020}, it was reported that
NUTS with Cayley transform is more efficient than the rejection
sampler of \citep{Hoff2009} in terms of the autocorrelations in the obtained
chains. Pourzanajani et al. \citep{Pourzanjani2021} applied NUTS
with Givens representation and GMC for uniform sampling on
the Stiefel manifold and network eigenmodel \citep{Hoff2009},
reporting the superiority of the former in terms of effective sample
size per iteration (ESS/iter). In \citep{Jauch2021}, three strategies,
namely, NUTS with polar expansion, GMC, and rejection
sampling were tested on the network eigenmodel, and it was shown that
NUTS with polar expansion is the best in terms of ESS/iter.
In \citep{Byrne2013}, GMC and rejection sampler were applied
to the network eigenmodel; the authors only showed that the latter
was stacked into local modes, whereas the former was not. The study proposing
NUTS with Householder parameterization \citep{Nirwan2019}
did not compare the proposed approach with other approaches. From those
previous studies, NUTS with some parameterization appears to be
better than the rejection sampler and GMC, whereas the order of NUTS with different parameterizations has not been investigated
rigorously. 

The primary objective of this study is to fill this gap by comparing
the numerical efficiency of the four parameterizations of orthogonal
matrices under identical conditions, and to explore the best current
practice. We apply NUTS with four alternative parameterizations
to four problems: uniform distribution, network eigenmodel, probabilistic
principal component analysis, and matrix completion for panel causal
analysis. The first three problems were adopted from the existing studies,
whereas the latter is novel to the literature. 

We evaluated the numerical efficiency of the different parameterizations
based on the minimum ESSs per second (minESS/sec) as well
as the minimum ESSs per iteration (minESS/iter). Our choice of
performance measure is notably different from that in the existing literature.
Although some studies have measured the numerical efficiency of posterior
simulator based on ESS/iter for certain parameters (\citep{Pourzanjani2021,Jauch2021}),
no existing study reports minESS/sec or minESS/iter. In the literature of
MCMC algorithms, it is a standard practice to evaluate the numerical
efficiency of a posterior simulator using minimum ESSs,
that is, the performance on the dimension that is the most difficult
to explore. In addition, even if minESS/iter is small, we can
effectively approximate the posterior distribution by running 
MCMC for longer duration because MCMC algorithms are justified based on an
asymptotic argument. Thus, for practitioners, minESS/sec is more important
than minESS/iter.

This study provides two takeaways. First, polar expansion appears
to be the best choice among the four parameterizations, particularly
with respect to minESS/sec. Although the other parameterizations work
for low-dimensional and/or simple problems, the relative advantage
of the polar expansion is evident in high-dimensional and/or difficult
problems. Second, when a sampling task is considerably high-dimensional
and/or difficult, none of the four parameterizations performed well.
This implies that the obtained posterior samples must be treated
with caution. This study highlighted the need for further improvements
in this area. 

The remainder of this paper is organized as follows. Section 2 introduces
the four alternative parameterizations. In Section 3, we apply 
NUTS with the four parameterizations to four statistical models, and
compare their numerical performance. Section 4 concludes the paper.

\section{parameterizations for Orthogonal Matrices}

A $J\times K$ orthogonal matrix $\boldsymbol{\Upsilon}$ is simulated
from the Stiefel manifold $\mathcal{V}^{J\times K}$, the set of $J\times K$
orthogonal matrices,

\[
\boldsymbol{\Upsilon}\in\mathcal{V}^{J\times K}=\left\{ \boldsymbol{\Upsilon}\in\mathbb{R}^{J\times K}:\boldsymbol{\Upsilon}^{\top}\boldsymbol{\Upsilon}=\boldsymbol{I}_{K}\right\} ,
\]
where $\boldsymbol{I}_{K}$ is a $K\times K$ identity matrix.
Let $\mathcal{D}$ denote the set of observations. The target kernel
is the posterior of $\boldsymbol{\Upsilon}$ conditional on $\mathcal{D}$
evaluated up to a normalizing constant, represented by 
product of the likelihood $f\left(\mathcal{D}|\boldsymbol{\Upsilon}\right)$
and prior density of $\boldsymbol{\Upsilon}$, $p\left(\boldsymbol{\Upsilon}\right)$,
$\pi\left(\boldsymbol{\Upsilon}\right)=f\left(\mathcal{D}|\boldsymbol{\Upsilon}\right)p\left(\boldsymbol{\Upsilon}\right)$.
For numerical efficiency, instead of dealing with $\boldsymbol{\Upsilon}$,
we parameterize $\boldsymbol{\Upsilon}$ using an auxiliary vector
$\boldsymbol{\varphi}$ and simulate $\boldsymbol{\varphi}$ from
an unconstrained space. The target kernel is modified by representing
$\boldsymbol{\Upsilon}$ as a matrix-valued function of $\boldsymbol{\varphi}$
and augmenting the determinant of the Jacobian of the transformation,
\[
\pi\left(\boldsymbol{\Upsilon}\left(\boldsymbol{\varphi}\right)\right)=f\left(\mathcal{D}|\boldsymbol{\Upsilon}\left(\boldsymbol{\varphi}\right)\right)p\left(\boldsymbol{\Upsilon}\left(\boldsymbol{\varphi}\right)\right)\left|\frac{\partial\boldsymbol{\Upsilon}}{\partial\boldsymbol{\varphi}^{\top}}\right|.
\]

In the following, we introduce four alternative parameterizations of
$\boldsymbol{\Upsilon}$. See \citep{Shepard2015} for a further discussion
on these parameterizations from mathematical and numerical points
of view.

\subsection{Polar expansion}

Jauch et al. \citep{Jauch2021} parameterized an orthogonal matrix
$\boldsymbol{\Upsilon}$ with a $J\times K$ unconstrained matrix
$\tilde{\boldsymbol{\Upsilon}}\in\mathbb{R}^{J\times K}$ using the
polar expansion, $\boldsymbol{\Upsilon}=\tilde{\boldsymbol{\Upsilon}}\left(\tilde{\boldsymbol{\Upsilon}}^{\top}\tilde{\boldsymbol{\Upsilon}}\right)^{-\frac{1}{2}}$,
where $\left(\tilde{\boldsymbol{\Upsilon}}^{\top}\tilde{\boldsymbol{\Upsilon}}\right)^{-\frac{1}{2}}$
denotes the inverse of the symmetric square root of $\tilde{\boldsymbol{\Upsilon}}^{\top}\tilde{\boldsymbol{\Upsilon}}$.
Representing the singular value decomposition of $\tilde{\boldsymbol{\Upsilon}}$
as $\tilde{\boldsymbol{\Upsilon}}=\boldsymbol{U}\boldsymbol{S}\boldsymbol{V}^{\top}$,
we get $\left(\tilde{\boldsymbol{\Upsilon}}^{\top}\tilde{\boldsymbol{\Upsilon}}\right)^{-\frac{1}{2}}=\boldsymbol{V}\boldsymbol{S}^{-\frac{1}{2}}\boldsymbol{V}^{\top}$.
We sample $\boldsymbol{\varphi}=\textrm{vec}\left(\tilde{\boldsymbol{\Upsilon}}\right)$
, where $\textrm{vec}\left(\cdot\right)$ denotes the column-wise
vectorization operator. Following \citep{Jauch2021}, we set the intermediate
distribution to the Wishart distribution. Then, the target kernel
is specified as
\[
\pi\left(\boldsymbol{\Upsilon}\left(\boldsymbol{\varphi}\right)\right)\propto f\left(\mathcal{D}|\boldsymbol{\Upsilon}\left(\boldsymbol{\varphi}\right)\right)p\left(\boldsymbol{\Upsilon}\left(\boldsymbol{\varphi}\right)\right)\exp\left(-\frac{1}{2}\boldsymbol{\varphi}^{\top}\boldsymbol{\varphi}\right).
\]

\subsection{Householder parameterization}

Nirwan and Bertschinger \citep{Nirwan2019} proposed a parameterization
based on Householder transformations. Orthogonal matrix $\boldsymbol{\Upsilon}$
is written as an ordered product of Householder reflectors:

\[
\boldsymbol{\Upsilon}=\boldsymbol{H}_{J}\left(\boldsymbol{v}_{J}\right)\boldsymbol{H}_{J-1}\left(\boldsymbol{v}_{J-1}\right)\cdots\boldsymbol{H}_{J-K+1}\left(\boldsymbol{v}_{J-K+1}\right)\boldsymbol{I}_{J\times K}
\]
\[
\boldsymbol{H}_{n}=\left(\begin{array}{cc}
\boldsymbol{I}_{J-n} & \boldsymbol{O}_{\left(J-n\right)\times n}\\
\boldsymbol{O}_{n\times\left(J-n\right)} & \tilde{\boldsymbol{H}}_{n}
\end{array}\right),
\]
\[
\quad\tilde{\boldsymbol{H}}_{n}=-\mathrm{sgn}\left(v_{n,1}\right)\left(\boldsymbol{I}_{J}-2\boldsymbol{u}_{n}\boldsymbol{u}_{n}^{\top}\right),
\]
\[
\boldsymbol{u}_{n}=\frac{\boldsymbol{v}_{n}+\mathrm{sgn}\left(v_{n,1}\right)\left\Vert \boldsymbol{v}_{n}\right\Vert \boldsymbol{e}_{n,1}}{\left\Vert \boldsymbol{v}_{n}+\mathrm{sgn}\left(v_{n,1}\right)\left\Vert \boldsymbol{v}_{n}\right\Vert \boldsymbol{e}_{n,1}\right\Vert },\quad\boldsymbol{v}_{n}\sim\mathcal{N}\left(\boldsymbol{0},\boldsymbol{I}\right),
\]
where $\boldsymbol{O}_{\left(J-K\right)\times K}$ is a $\left(J-K\right)\times K$
matrix of zeros, $\boldsymbol{I}_{J\times K}$ is a $J\times K$
matrix with the identity matrix as its top block and the remaining
entries zero, $\boldsymbol{e}_{n,1}=\left(1,\boldsymbol{0}_{n-1}^{\top}\right)^{\top}$,
$\mathrm{sgn}\left(\cdot\right)$ is the sign operator, and $\left\Vert \cdot\right\Vert $
is the Euclidean norm. The vector to be sampled is $\boldsymbol{\varphi}=\left(\boldsymbol{v}_{J}^{\top},\boldsymbol{v}_{J-1}^{\top},...,\boldsymbol{v}_{J-K+1}^{\top}\right)^{\top}$.
Using this approach, we can avoid computing the Jacobian adjustment term.

\subsection{Cayley transform}

Jauch et al. \citep{Jauch2020} parameterized orthogonal matrices using
the modified Cayley transform \citep{Shepard2015}. Given a $K\times K$
skew symmetric matrix 
\[
\boldsymbol{X}\in\mathrm{Skew}\left(J\right)=\left\{ \boldsymbol{X}\in\mathbb{R}^{J\times J}:\boldsymbol{X}=-\boldsymbol{X}^{\top}\right\} ,
\]
the modified Cayley transform of $\boldsymbol{X}$ can be written as
\[
\boldsymbol{\Upsilon}=\left(\boldsymbol{I}_{J}+\boldsymbol{X}\right)\left(\boldsymbol{I}_{J}-\boldsymbol{X}\right)^{-1}\boldsymbol{I}_{J\times K}.
\]
For brevity, hereinafter we call this type of transformation as Cayley transform.
$\boldsymbol{X}$ has the following block structure:
\[
\boldsymbol{X}=\left(\begin{array}{cc}
\boldsymbol{B} & -\boldsymbol{A}^{\top}\\
\boldsymbol{A} & \boldsymbol{O}_{\left(J-K\right)\times\left(J-K\right)}
\end{array}\right),
\]
where $\boldsymbol{A}\in\mathbb{R}^{\left(J-K\right)\times K}$ and
$\boldsymbol{B}\in\mathrm{Skew}\left(K\right)$. Let $\boldsymbol{b}$
be a $K\left(K-1\right)/2$-dimensional vector of independent elements
of $\boldsymbol{B}$ obtained by eliminating the diagonal and supradiagonal
elements of $\textrm{vec}\left(\boldsymbol{B}\right)$. We sample
$\boldsymbol{\varphi}=\left(\boldsymbol{b}^{\top},\textrm{vec}\left(\boldsymbol{A}^{\top}\right)\right)^{\top}$.
The Jacobian adjustment term is defined as follows:
\[
\left|\frac{\partial\boldsymbol{\Upsilon}}{\partial\boldsymbol{\varphi}^{\top}}\right|=\left|2^{2}\boldsymbol{\Gamma}^{\top}\left(\boldsymbol{G}_{1}\otimes\boldsymbol{G}_{2}\right)\boldsymbol{\Gamma}\right|^{\frac{1}{2}},
\]
\[
\boldsymbol{G}_{1}=\left(\boldsymbol{I}_{J}-\boldsymbol{X}\right)^{-1}\boldsymbol{I}_{J\times K}\boldsymbol{I}_{J\times K}^{\top}\left(\boldsymbol{I}_{J}-\boldsymbol{X}\right)^{-\top},
\]
\[
\quad\boldsymbol{G}_{2}=\left(\boldsymbol{I}_{J}-\boldsymbol{X}\right)^{-\top}\left(\boldsymbol{I}_{J}-\boldsymbol{X}\right)^{-1},
\]
\[
\boldsymbol{\Gamma}=\left(\begin{array}{cc}
\boldsymbol{\Gamma}_{1} & \boldsymbol{\Gamma}_{2}\end{array}\right),
\]
\[
\boldsymbol{\Gamma}_{1}=\left(\boldsymbol{\Theta}_{1}^{\top}\otimes\boldsymbol{\Theta}_{1}^{\top}\right)\boldsymbol{D}_{K},\quad\boldsymbol{\Gamma}_{2}=\left(\boldsymbol{I}_{J^{2}}-\boldsymbol{K}_{J,J}\right)\left(\boldsymbol{\Theta}_{1}^{\top}\otimes\boldsymbol{\Theta}_{2}^{\top}\right),
\]
\[
\boldsymbol{\Theta}_{1}=\left(\begin{array}{cc}
\boldsymbol{I}_{K} & \boldsymbol{O}_{K\times\left(J-K\right)}\end{array}\right),\quad\boldsymbol{\Theta}_{2}=\left(\begin{array}{cc}
\boldsymbol{O}_{\left(J-K\right)\times K} & \boldsymbol{I}_{J-K}\end{array}\right),
\]
where $\boldsymbol{D}_{K}$ is a $K^{2}\times K\left(K-1\right)/2$
matrix such that $\boldsymbol{D}_{K}\boldsymbol{b}=\mathrm{vec}\left(\boldsymbol{B}\right)$ and 
$\boldsymbol{K}_{J,J}$ is the commutation matrix that satisfies $\boldsymbol{K}_{J,J}\mathrm{vec}\left(\boldsymbol{A}\right)=\mathrm{vec}\left(\boldsymbol{A}\right)^{\top}$.
See Appendix B of \citep{Jauch2020} (pp. 1581-1582) for further details
to construct $\boldsymbol{D}_{K}$ and $\boldsymbol{K}_{J,J}$.

\subsection{Givens representation}

Pourzanjani et al. \citep{Pourzanjani2021} proposed to use a Givens
representation of $\boldsymbol{\Upsilon}$:
\begin{eqnarray*}
\boldsymbol{\Upsilon} & = & \boldsymbol{R}_{1,2}\left(\theta_{1,2}\right)\cdots\boldsymbol{R}_{1,J}\left(\theta_{1,J}\right)\cdots\boldsymbol{R}_{2,3}\left(\theta_{2,3}\right)\cdots\boldsymbol{R}_{2,J}\left(\theta_{2,J}\right)\\
 &  & \cdots\boldsymbol{R}_{K,K+1}\left(\theta_{K,K+1}\right)\boldsymbol{R}_{K,J}\left(\theta_{K,J}\right)\boldsymbol{I}_{J\times K},
\end{eqnarray*}
where $\theta_{k,j}$ is a coordinate variable, $\boldsymbol{R}_{k,j}\left(\theta_{k,j}\right)$
is a $J\times J$ matrix that takes the form of an identity matrix
except for the $\left(j,j\right)$ and $\left(k,k\right)$ positions
which are replaced by $\cos\theta_{j,k}$ and the $\left(j,k\right)$
and $\left(k,j\right)$ positions which are replaced by $-\sin\theta_{j,k}$
and $\sin\theta_{k,j}$, respectively. $\theta_{1,2},\theta_{2,3},...,\theta_{K,K+1}\in\left(-\pi,\pi\right)$
and the remaining coordinates are in the range $\left(-2\pi,2\pi\right)$. This
parameterization can induce the posterior of $\theta_{j,k}$ to be multimodal
\citep{Pourzanjani2021}. To address this, Pourzanjani et
al. \citep{Pourzanjani2021} suggested to further reparameterize $\theta_{k,j}$ with
an independent auxiliary parameter $r_{k,j}$,
\[
\varphi_{k,j}^{\flat}=r_{k,j}\cos\theta_{k,j},\quad\varphi_{k,j}^{\sharp}=r_{k,j}\sin\theta_{k,j}.
\]
The marginal distribution of $r_{k,j}$ is given by a normal distribution
with a mean of one and a standard deviation of 0.1, $r_{k,j}\sim\mathcal{N}\left(1,0.1^{2}\right)$. 
See Section 4.2 of \citep{Pourzanjani2021} (pp. 647-653) for further
discussion. The Jacobian adjustment term is:
\[
g\left(\boldsymbol{\varphi}\right)=\prod_{j=1}^{J}\prod_{k=j+1}^{K}\left(\cos\theta_{k,j}\right)^{k-j-1}.
\]

Table 1 summarizes the number of essential parameters, i.e.,
the dimension of $\boldsymbol{\varphi}$, for the four alternative parameterizations.
The polar expansion has the largest number of essential parameters,
whereas the Householder and Cayley transforms the
smallest. 

\begin{table}
\caption{Number of essential parameters}

\medskip{}

\centering{}%
\begin{tabular}{lc}
\hline 
Approach & \# of essential parameters\tabularnewline
\hline 
Polar & $JK$\tabularnewline
Householder & $JK-K\left(K-1\right)/2$\tabularnewline
Cayley & $K\left(K-1\right)/2+J\left(J-K\right)$\tabularnewline
Givens & $JK-K\left(K-1\right)/2$\tabularnewline
\hline 
\end{tabular}
\end{table}

\section{Comparative Simulation Study}

This section compares the numerical efficiencies of the alternative parameterizations
on four testing benches: uniform distribution, network eigenmodel,
probabilistic principal component analysis, and matrix completion
for panel causal analysis. We employed NUTS \citep{Hoffman2014},
an adaptive version of the Hamiltonian Monte Carlo \citep{Duane1987,Neal2011}.
All computations were performed using CmdStan software (version 2.33.1)
\footnote{https://mc-stan.org/users/interfaces/cmdstan} with R (version
4.1.2) \citep{RCT2021}\footnote{https://cran.r-project.org/} and
\texttt{cmdstanr} package (version 0.7.0), running on an Ubuntu 22.04.3
desktop with AMD Ryzen Threadripper 3990X 2.9GHz processor. All the
stan files were based on the replication codes provided by the authors
of the original papers.\footnote{The replication codes are available from the following websites.

Householder parameterization:

$\quad$https://github.com/RSNirwan/HouseholderBPCA

Cayley transform:

$\quad$https://github.com/michaeljauch/cayley

Polar expansion:

$\quad$https://github.com/michaeljauch/polar

Givens representation:

$\quad$https://github.com/pourzanj/TfRotationPca/tree/master} For each posterior simulation, we generated 1,000 posterior draws
and used the final 500 draws for evaluation. The numerical efficiency
of each approach was measured using minESS for the main iterations, obtained using the \texttt{mcmcse} package
(version 1.5-0). minESS was normalized by the number of iterations and
wall-clock elapsed time, denoted as minESS/iter and minESS/sec, respectively.
We report the averages of 64 runs.

\subsection{Uniform sampling on the Stiefel manifold}

We compared the alternative parameterizations for uniform sampling
on the Stiefel manifold. The dimension of the target distribution
was set to combinations of $K\in\left\{ 3,10\right\} $ and $J\in\left\{ 10,100,200\right\} $.
We did not examine an approach based on the Cayley transform for these situations
with $J=K$ because it requires considerable coding effort.\footnote{When $J=K$, $\boldsymbol{A}$ disappears and it holds that $\boldsymbol{X}=\boldsymbol{B}$.
Thus, conditional executions must be added to virtually all lines
involving $\boldsymbol{X}$. } Owing to the limitations of the computational budget, we did not test the
Givens representation for the most high-dimensional case, $\left(J,K\right)=\left(200,10\right)$.

Table 2 presents the simulation results. Polar expansion was the
best parameterization regardless of the efficiency measure. The win margin was larger for
high-dimensional cases. For the minESS/iter metric, the remaining three
parameterizations were comparable. For minESS/sec, Householder
was the most efficient parameterization after polar expansion, whereas the Givens
representation was the worst. 

\begin{table}
\caption{Simulation result: Uniform sampling}

\medskip{}

\centering{}%
\begin{tabular}{rrrrrr}
\multicolumn{6}{l}{(a) minESS/iter}\tabularnewline
\hline 
$J$ & $K$ & Polar & Householder & Cayley & Givens\tabularnewline
\hline 
10 & 3 & \textbf{1.005} & 0.403 & 0.225 & 0.296\tabularnewline
100 & 3 & \textbf{0.972} & 0.220 & 0.202 & 0.244\tabularnewline
200 & 3 & \textbf{0.663} & 0.226 & 0.177 & --\tabularnewline
\hline 
10 & 10 & \textbf{0.938} & 0.485 & -- & 0.242\tabularnewline
100 & 10 & \textbf{0.465} & 0.270 & 0.163 & 0.175\tabularnewline
200 & 10 & \textbf{0.384} & 0.204 & 0.136 & --\tabularnewline
\hline 
 &  &  &  &  & \tabularnewline
\multicolumn{6}{l}{(b) minESS/sec}\tabularnewline
\hline 
$J$ & $K$ & Polar & Householder & Cayley & Givens\tabularnewline
\hline 
10 & 3 & \textbf{11137.636} & 2420.666 & 225.096 & 18.020\tabularnewline
100 & 3 & \textbf{1360.927} & 6.774 & 0.399 & 0.016\tabularnewline
200 & 3 & \textbf{481.104} & 0.820 & 0.061 & --\tabularnewline
\hline 
10 & 10 & \textbf{1804.136} & 965.346 & -- & 3.363\tabularnewline
100 & 10 & \textbf{185.496} & 1.671 & 0.019 & 0.001\tabularnewline
200 & 10 & \textbf{64.135} & 0.134 & 0.001 & --\tabularnewline
\hline 
\end{tabular}
\end{table}

\subsection{Network eigenmodel}

The network eigenmodel was first introduced by \citet{Hoff2009}
and has been widely used as a testing bench \citep{Byrne2013,Jauch2021,Pourzanjani2021}; however,
a thorough comparison of alternative parameterizations
is absent. A graph matrix represents how proteins in a protein
network interact with each other. Probability of a connection
between proteins is modeled using a symmetric matrix of probabilities
whose rank is much smaller than the dimension of the observed graph matrix. Given
a $J\times J$ symmetric graph matrix $\boldsymbol{Y}=\left(y_{j,j^{\prime}}\right)$
with $y_{j,j^{\prime}}\in\left\{ 0,1\right\} $, the probability of
the connections is specified as follows: For $j,j^{\prime}=1,...,J$,
\[
y_{j,j^{\prime}}\sim\mathrm{Bernoulli}\left(\Phi\left(\left(\boldsymbol{\Upsilon}\boldsymbol{\Lambda}\boldsymbol{\Upsilon}^{\top}\right)_{j,j^{\prime}}+\mu\right)\right),
\]
where $\boldsymbol{\Lambda}=\mathrm{diag}\left(\lambda_{1},\lambda_{2},\lambda_{3}\right)$,
$\boldsymbol{\Upsilon}\in\mathcal{V}^{J\times K}$, $\mu$, and $\lambda_{k}$
are unknown parameters, and $\Phi\left(\cdot\right)$ is the probit
link function. Following previous studies, we selected $K=3$. We
assign normal priors to $\mu$ and $\lambda_{k}$: $\mu\sim\mathcal{N}\left(0,10^{2}\right)$
$\lambda_{k}\sim\mathcal{N}\left(0,J\right)$. The dimension of
the dataset was $J=270$. 

As shown in Table 3, none of the four parameterizations performed
well. Polar expansion was the best, but its minESS/iter and minESS/sec were too small to conduct a reliable
posterior analysis within a reasonable time. Previous studies reported
minESS/iter for only some of the unknown parameters. Although not
reported in Table 3, we confirmed the results of \citep{Jauch2021}
and \citep{Pourzanjani2021}. As reported in these papers, some of
the unknown parameters (e.g., $\lambda_{k}$s and $c$) are effectively
simulated, but the posterior draws of some entries in $\boldsymbol{\Upsilon}$
are strongly autocorrelated. 

\begin{table}
\caption{Simulation result: Network eigenmodel}

\medskip{}

\centering{}%
\begin{tabular}{rrrrrr}
\multicolumn{6}{l}{(a) minESS/iter}\tabularnewline
\hline 
$J$ & $K$ & Polar & Householder & Cayley & Givens\tabularnewline
\hline 
270 & 3 & \textbf{0.018} & 0.012 & 0.009 & 0.009\tabularnewline
\hline 
 &  &  &  &  & \tabularnewline
\multicolumn{6}{l}{(b) minESS/sec}\tabularnewline
\hline 
$J$ & $K$ & Polar & Householder & Cayley & Givens\tabularnewline
\hline 
270 & 3 & \textbf{0.075} & 0.000 & 0.001 & 0.001\tabularnewline
\hline 
\end{tabular}
\end{table}

\subsection{Probabilistic principal component analysis}

In probabilistic principal component analysis \citep{Tipping2002},
a $J$-dimensional demeaned vector $\boldsymbol{y}_{i}\in\mathbb{R}^{J}$
was modeled using a linear function of low-dimensional latent state $\boldsymbol{z}_{i}\in\mathbb{R}^{K}$
with $K<J$. The distribution of $\boldsymbol{y}_{i}$ is specified
as follows: For $i=1,...,N$,
\[
\boldsymbol{x}_{i}|\boldsymbol{z}_{i}\sim\mathcal{N}\left(\boldsymbol{W}\boldsymbol{\Lambda}\boldsymbol{z}_{i},\;\sigma^{2}\boldsymbol{I}_{J}\right),\quad\boldsymbol{z}_{i}\sim\mathcal{N}\left(\boldsymbol{0}_{K},\boldsymbol{I}_{K}\right),
\]
where $\boldsymbol{\Lambda}=\mathrm{diag}\left(\lambda_{1},...,\lambda_{K}\right)$,
$\boldsymbol{W}\in\mathcal{V}^{J\times K}$, and $\lambda_{k}$ are
unknown parameters, and $\sigma^{2}$ is the variance of the idiosyncratic
shocks. This model is closely related to static latent factor modeling
(see \citep{Lopes2014} for an overview). $\boldsymbol{z}_{i}$ can be seen as a vector of latent factors
and the quantity $\boldsymbol{W}\boldsymbol{\Lambda}$ as a factor
loading matrix.

We applied this model to two specifications of synthetic data. The first specification
was adopted from \citep{Nirwan2019}: $N=150$,
$J=5$, $K=2$, $\left(\lambda_{1},\lambda_{2}\right)=\left(9,1\right)$,
and $\sigma^{2}=0.01^{2}$\emph{. }The second\emph{ } specification was adopted
from \citep{Jauch2020,Pourzanjani2021}: $N=100$, $J=50$, $K=3$,
$\left(\lambda_{1},\lambda_{2},\lambda_{3}\right)=\left(5,3,1.5\right)$,
and $\sigma^{2}=1$. For both specifications, $\boldsymbol{W}$
was generated uniformly from $\mathcal{V}^{50\times3}$\footnote{First, each entry of $\boldsymbol{A}\in\mathbb{R}^{J\times K}$ is
simulated independently from the standard normal distribution and
then $\boldsymbol{W}$ is obtained through a polar expansion, $\boldsymbol{W}=\boldsymbol{A}\left(\boldsymbol{A}^{\top}\boldsymbol{A}\right)^{-\frac{1}{2}}$
(see Proposition 7.1 of \citep{Eaton1989}, pp. 100-101).}. Non-informative priors were employed for all the parameters and cases. 

The first two rows of panels (a) and (b) in Table 4 summarize the
results. In terms of minESS/iter, the performance order of
the four parameterizations are not clear. For the first dataset, polar expansion performed the best. For the second dataset, Cayley transform performed the best. When the performance
is evaluated based on the minESS/sec metric, for the first dataset, 
polar expansion and Cayley transformation are largely comparable and
superior to the other parameterizations. For the second dataset, polar
expansion performed the best. Givens representation did not
work well for either dataset. Similar to uniform sampling
cases, in terms of minESS/iter, Givens representation
was marginally inferior to Householder and Cayley transformations, while Givens representation was
slow to draw, leading to a much smaller minESS/sec. 

\begin{table}
\caption{Simulation result: Probabilistic principal component analysis}

\medskip{}

\centering{}%
\begin{tabular}{lrrrrrr}
\multicolumn{7}{l}{(a) minESS/iter}\tabularnewline
\hline 
 & $J$ & $K$ & Polar & Householder & Cayley & Givens\tabularnewline
\hline 
Synthetic 1 & 5 & 2 & \textbf{0.819} & 0.120 & 0.249 & 0.100\tabularnewline
Synthetic 2 & 50 & 3 & 0.077 & 0.043 & \textbf{0.131} & 0.056\tabularnewline
\hline 
\multirow{1}{*}{Real} & 569 & 2 & \textbf{0.240} & 0.027 & 0.046 & --\tabularnewline
\hline 
\multicolumn{2}{l}{} & \tabularnewline
\multicolumn{7}{l}{(b) minESS/sec}\tabularnewline
\hline 
 & $J$ & $K$ & Polar & Householder & Cayley & Givens\tabularnewline
\hline 
Synthetic 1 & 5 & 2 & 240.253 & 131.043 & \textbf{262.117} & 30.558\tabularnewline
Synthetic 2 & 50 & 3 & \textbf{11.828} & 0.131 & 1.784 & 0.009\tabularnewline
\hline 
\multirow{1}{*}{Real} & 569 & 2 & \textbf{2.280} & 0.509 & 0.206 & --\tabularnewline
\hline 
\end{tabular}
\end{table}

Following \citep{Nirwan2019}, we applied the model to
the breast cancer Wisconsin dataset retrieved from scikit-learn,
a machine learning library for the Python programming language \citep{Pedregosa2011}.
The model was extended by incorporating a mean vector $\boldsymbol{\mu}$
as 
\[
\boldsymbol{y}_{i}|\boldsymbol{z}_{i}\sim\mathcal{N}\left(\boldsymbol{\mu}+\boldsymbol{W}\boldsymbol{\Lambda}\boldsymbol{z}_{i},\;\sigma^{2}\boldsymbol{I}_{J}\right).
\]
We assigned a non-informative prior to $\boldsymbol{\mu}$: $p\left(\boldsymbol{\mu}\right)\propto1$.
As shown in Table 4, polar expansion performed the best, irrespective
of the performance measure. We do not report on Givens representation
because the posterior simulation was intolerably slow.

\subsection{Matrix completion for causal analysis}

The final testing bench is matrix completion for causal analysis.
Matrix completion has been studied extensively in the machine learning
community (e.g., \citep{Salakhutdinov2007,Ding2011,Babacan2012,Mai2015,Farias2022,Tanaka2022,Yuchi2023,Zhai2023}).
It has been applied to causal analysis with event study designs
using panel data (e.g., \citep{Samartsidis2020,Nethery2021,Tanaka2021,Pang2022,Samartsidisforthcoming}).
In this framework, outcomes under treatment are removed from an outcome
matrix, and the outcome matrix is ``completed'', treating the missing
entries as unrealized potential outcomes under treatment, or counterfactual
outcomes. By comparing between the estimates of the counterfactual
and realized outcomes, we can infer the treatment effects. 

Let $\boldsymbol{Y}\in\mathbb{R}^{J\times T}$ denote the matrix of
outcomes for $J$ entities and $T$ time periods. The elements in $\boldsymbol{Y}$
corresponding to the treated entities and the time periods are treated as
missing data. We can infer unrealized potential outcomes by completing
$\boldsymbol{Y}$. Let $\boldsymbol{Y}^{\mathrm{miss}}$ denote the
set of missing entries in $\boldsymbol{Y}$ and $\boldsymbol{Y}^{\mathrm{obs}}$
denote the set of observed entries. 

Given $\boldsymbol{Y}^{\mathrm{miss}}$, the model of ``completed''
$\boldsymbol{Y}$ is composed of three matrices, $\boldsymbol{Y}=\boldsymbol{\Xi}+\boldsymbol{\Gamma}+\boldsymbol{U}$.
First, $\boldsymbol{\Xi}\in\mathbb{R}^{J\times T}$ is a matrix of
covariate effects, $\boldsymbol{x}_{j,t}$ is a vector of covariates,
$\boldsymbol{\Xi}=\left(\xi_{j,t}\right)$ with $\xi_{j,t}=\boldsymbol{\beta}^{\top}\boldsymbol{x}_{j,t}$.
Second, $\boldsymbol{\Gamma}\in\mathbb{R}^{J\times T}$ is a matrix
with rank $K<\min\left(J,T\right)$ that is factorized as in singular
value decomposition, $\boldsymbol{\Gamma}=\boldsymbol{\Phi}\boldsymbol{\Lambda}\boldsymbol{\Psi}^{\top}$,
where $\boldsymbol{\Lambda}=\mathrm{diag}\left(\lambda_{1},...,\lambda_{K}\right)$
is a diagonal matrix, $\boldsymbol{\Phi}=\left(\phi_{j,k}\right)\in\mathcal{V}^{J\times K}$
and $\boldsymbol{\Psi}=\left(\psi_{t,k}\right)\in\mathcal{V}^{T\times K}$
are orthogonal matrices. Third, $\boldsymbol{U}=\left(u_{j,k}\right)\in\mathbb{R}^{J\times T}$
is a matrix of error terms whose entries are distributed according
to an independent normal distribution with variance $\sigma^{2}$,
$u_{j,t}\sim\mathcal{N}\left(0,\sigma^{2}\right).$ 

We conducted a posterior simulation treating $\boldsymbol{Y}^{\mathrm{miss}}$
as unknown parameters. Let $\boldsymbol{X}$ denote the set of covariates.
Then, the likelihood of the ``completed'' $\boldsymbol{Y}$ has
the standard form:
\[
p\left(\boldsymbol{Y}|\boldsymbol{Y}^{\mathrm{miss}},\boldsymbol{\Phi},\boldsymbol{\Lambda},\boldsymbol{\Psi},\boldsymbol{\beta},\sigma^{2};\boldsymbol{X}\right)=
\]
\[
\left(2\pi\sigma^{2}\right)^{-\frac{JT}{2}}\exp\left\{ -\frac{1}{2\sigma^{2}}\left\Vert \boldsymbol{Y}-\boldsymbol{\Phi}\boldsymbol{\Lambda}\boldsymbol{\Psi}^{\top}-\boldsymbol{\Xi}\right\Vert _{F}^{2}\right\} .
\]
The target kernel is represented as:
\[
p\left(\boldsymbol{Y}^{\mathrm{miss}},\boldsymbol{\Phi},\boldsymbol{\Lambda},\boldsymbol{\Psi},\boldsymbol{\beta},\sigma^{2}|\boldsymbol{Y}^{\mathrm{obs}},\boldsymbol{X}\right)\propto 
\]
\[
p\left(\boldsymbol{Y}|\boldsymbol{Y}^{\mathrm{miss}},\boldsymbol{\Phi},\boldsymbol{\Lambda},\boldsymbol{\Psi},\boldsymbol{\beta},\sigma^{2};\boldsymbol{X}\right)p\left(\boldsymbol{Y}^{\mathrm{miss}},\boldsymbol{\Phi},\boldsymbol{\Lambda},\boldsymbol{\Psi},\boldsymbol{\beta},\sigma^{2}\right)
\]
where $p\left(\boldsymbol{Y}^{\mathrm{miss}},\boldsymbol{\Phi},\boldsymbol{\Lambda},\boldsymbol{\Psi},\boldsymbol{\beta},\sigma^{2}\right)$
is a prior. For $\lambda_{k}$, we employed an exponential prior $\lambda_{k}\sim\mathcal{E}\left(\eta\right)$,
where $\eta$ is a prefixed rate parameter. This choice can be seen
as a Bayesian counterpart to matrix completion with the nuclear norm
penalty \citep{Du2023}. We assigned uniform Haar prior to $\boldsymbol{\Phi}$
and $\boldsymbol{\Psi}$: $p\left(\boldsymbol{\Phi}\right)\propto1$
and $p\left(\boldsymbol{\Psi}\right)\propto1$, non-informative prior
to $\boldsymbol{Y}^{\mathrm{miss}}$ and $\boldsymbol{\beta}$: $p\left(\boldsymbol{Y}^{\mathrm{miss}}\right)\propto1$,
$p\left(\boldsymbol{\beta}\right)\propto1$, and Jeffreys prior
to $\sigma^{2}$: $p\left(\sigma^{2}\right)\propto\sigma^{-2}$. 

Using this model, we re-examined the empirical analysis of carbon taxes
on $\mathrm{CO}_{2}$ emissions in \citep{Andersson2019} which uses
synthetic control method \citep{Abadie2003,Abadie2010}. We used the annual
panel data on per capita $\mathrm{CO}_{2}$ emissions from transport,
covering the years 1960-2005 for 14 OECD (Organisation for Economic
Co-operation and Development) countries: Australia, Belgium, Canada,
Denmark, France, Greece, Iceland, Japan, New Zealand, Poland, Portugal,
Spain, Sweden, Switzerland, and the United States. Thus, $J=14$
and $T=46$. Sweden implemented carbon taxes in 1990, whereas the others
did not. We exploited this event as a quasi-experiment to infer
the effects of the carbon taxes on $\mathrm{CO}_{2}$ emissions in
Sweden, treating the other countries as a control group. See \citep{Andersson2019}
for further details on the dataset and a discussion on the choice
of the control group. Three cases with $K\in\left\{ 3,5,10\right\} $
were considered.

Table 5 presents the results. Using Givens representation, for
some runs with $K=10$, we obtained extremely autocorrelated chains
and could not effectively compute the ESS due to numerical
instabilities. Based on the minESS/iter metric, the performance of the
four parameterizations were similarly poor. For the minESS/sec metric, polar expansion ranked the best.

\begin{table}
\caption{Simulation result: Matrix completion for causal analysis }

\medskip{}

\centering{}%
\begin{tabular}{rrrrr}
\multicolumn{5}{l}{(a) minESS/iter}\tabularnewline
\hline 
$K$ & Polar & Householder & Cayley & Givens\tabularnewline
\hline 
3 & 0.027 & \textbf{0.027} & 0.027 & 0.027\tabularnewline
5 & 0.028 & 0.027 & \textbf{0.028} & 0.027\tabularnewline
10 & 0.027 & \textbf{0.027} & 0.027 & --\tabularnewline
\hline 
\multicolumn{5}{l}{}\tabularnewline
\multicolumn{5}{l}{(b) minESS/sec}\tabularnewline
\hline 
$K$ & Polar & Householder & Cayley & Givens\tabularnewline
\hline 
3 & \textbf{4.047} & 1.446 & 0.032 & 0.012\tabularnewline
5 & \textbf{1.795} & 0.766 & 0.005 & 0.003\tabularnewline
10 & \textbf{0.764} & 0.047 & 0.004 & --\tabularnewline
\hline 
\end{tabular}
\end{table}

\section{Conclusions}

To determine the best practice for Monte Carlo simulation
on the Stiefel manifold, we compared four parameterizations of orthogonal
matrices, namely, polar expansion, Householder transformation,
 (modified) Cayley transformation, and Givens representation,
for various statistical applications, and compared their numerical performance
based on the effective sample size. Series of simulations using NUTS revealed that polar expansion is the best among
the four parameterizations. However, when the sampling space is high-dimensional
and/or complex, all four approaches are unlikely to work 
well. Although the poor quality of a sampler does not break the theoretical
(asymptotic) validity of a posterior simulation, it is necessary to
generate longer chains and carefully examine the obtained draws,
particularly when the sampling problem is high-dimensional and/or difficult.
Therefore, further research is required in this area.

\begin{acks}
This study was supported by JSPS KAKENHI Grant Number 20K22096.
\end{acks}

\bibliographystyle{ACM-Reference-Format}
\nocite{*}
\bibliography{reference}

\end{document}